\documentclass[final,5p,twocolumn]{elsarticle}

\usepackage{graphicx}

\usepackage{amssymb}
\usepackage{amsmath}

\biboptions{sort&compress}

\newcommand{\beq}{\begin{equation}}
\newcommand{\eeq}{\end{equation}}

\begin{document}

\begin{frontmatter}



\title{Matrix algorithms for solving (in)homogeneous bound state equations}


\author{M.~Blank, A.~Krassnigg}
\ead{martina.blank@uni-graz.at~andreas.krassnigg@uni-graz.at}
\address{Institut f\"ur Physik, Universit\"at Graz,\\ Universit\"atsplatz 5, 8010 Graz, Austria}

\begin{abstract}
In the functional approach to quantum chromodynamics, the properties of hadronic bound states
are accessible via covariant integral equations, e.g.~the Bethe-Salpeter equations for mesons. In
particular, one has to deal with linear, homogeneous integral equations which, in sophisticated model
setups, use numerical representations of the solutions of other integral equations as part of their input.
Analogously, inhomogeneous equations can be constructed to obtain off-shell information in addition to
bound-state masses and other properties obtained from the covariant analogue to a wave function of the
bound state. These can be solved very efficiently using well-known matrix algorithms for eigenvalues
(in the homogeneous case) and the solution of linear systems (in the inhomogeneous case). We demonstrate
this by solving the homogeneous and inhomogeneous Bethe-Salpeter equations and find, e.g.~that for the
calculation of the mass spectrum it is more efficient to use the inhomogeneous equation. This is valuable
insight, in particular for the study of baryons in a three-quark setup and more involved systems.
\end{abstract}


\end{frontmatter}


\section{Introduction}
The underlying quantum field theory of the strong interaction in the standard model of elementary particle physics is
quantum chromodynamics (QCD), a non-abelian gauge theory which deals with elementary degrees of freedom
called quarks and gluons \cite{Marciano:1977su}. A remarkable feature of QCD is asymptotic freedom, which means that
the running coupling of the theory is small in the high-energy regime \cite{Gross:1973id,Gross:1973ju,Politzer:1973fx}.
There, perturbation theory can be applied, and perturbative QCD has been well established in the high-energy domain,
(e.\,g.~\cite{Lepage:1980fj} and references therein). At low energies, however, perturbation theory is no longer applicable,
since the value of the running coupling increases to the order of 1. Since bound states are intrinsically nonperturbative,
corresponding methods have been developed and used to investigate hadrons, the bound states of quarks and gluons.
We eclectically list a few references regarding constituent quark models
\cite{Godfrey:1985xj,Capstick:1986bm,Glozman:1998ag,Barnes:2005pb,Melde:2008yr},
effective field theories \cite{Gasser:1983yg,Scherer:2002tk,Brambilla:2004wf},
lattice-regularized QCD \cite{McNeile:2006nv,Burch:2006dg,Wada:2007cp,Dudek:2007wv,Gregory:2008mn,Gattringer:2010zz},
QCD sum rules \cite{Colangelo:2000dp,Jamin:2001fw,Maltman:2001gc,Penin:2001ux,Ball:2004ye,Lucha:2009jk},
and the renormalization-group approach to QCD \cite{Gies:2001nw,Pawlowski:2005xe}
(always see also references therein).

Another remarkable property closely related to bound states is the so-called confinement of quarks and gluons. 
It entails that only objects like hadrons, where the color charges carried by the elementary degrees of freedom 
are combined to a color-neutral state, can be observed directly. While in constituent-quark models confinement is 
usually implemented via potential terms of an infinitely rising nature (of, e.g., harmonic-oscillator or linear type), 
in QCD the particularities are more delicate (for a recent review of the problems surrounding quark confinement, 
see e.g.~\cite{Alkofer:2006fu}). In a quantum field theoretical setup,
as we use it here, confinement is tied to the properties of the fundamental Green functions of the theory.

In the present work, we employ the Dyson-Schwinger-equation (DSE) approach to QCD. The DSEs are the covariant and nonperturbative
continuum equations of motion in quantum field theory. They constitute an infinite set of coupled and in general nonlinear
integral equations for the Green functions of the quantum field theory under consideration. There are several extensive reviews
on the subject that focus on different aspects of DSEs, like fundamental Green functions \cite{Roberts:1994dr,Alkofer:2000wg,Fischer:2006ub},
bound-state calculations \cite{Maris:2003vk,Roberts:2007jh} and applications of the formalism, e.g.~to QCD at finite temperature and
density \cite{Roberts:2000aa}. Bound states are studied in this approach with the help of covariant equations embedded in
the system of DSEs. In particular, the Bethe-Salpeter equation (BSE) \cite{Bethe:1951bs,Salpeter:1951sz} is used for 
two-body problems such as mesons \cite{Smith:1969az,Jain:1993qh,Krassnigg:2009zh} and covariant
Faddeev-type equations \cite{Faddeev:1960su} are used for three-body problems such as baryons \cite{Eichmann:2009qa}.

Ideally, one could obtain a self-consistent simultaneous solution of all DSEs, which would be equivalent to
a solution of the underlying quantum field theory. While in investigations of certain aspects of the theory
such an approach is successful (see, e.\,g.~\cite{Alkofer:2008tt,Fischer:2008uz} and references therein),
numerical studies of hadrons necessitate a truncation of this infinite tower of equations.

\begin{figure*}
\begin{center}
\includegraphics[scale=0.9,clip=true]{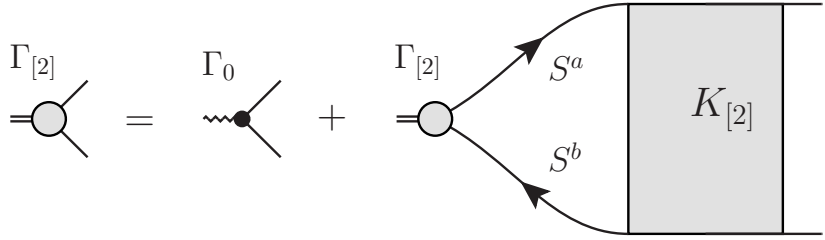}
\caption{\label{fig:inhom_bse} The inhomogeneous vertex BSE, Eq.~(\ref{eq:inhom_bse})}
\end{center}
\end{figure*}
Once the covariant bound-state equation has been solved to obtain the mass spectrum of the system under
consideration, the corresponding covariant amplitudes can be used to compute further observables. In
rainbow-ladder truncation, prominent examples include
leptonic decay constants \cite{Maris:1997tm,Maris:1999nt,Ivanov:1998ms}, hadronic decays \cite{Jarecke:2002xd,Jarecke:2005ph},
and electromagnetic properties of both mesons \cite{Ivanov:1997iu,Maris:1999bh,Maris:2000sk,Holl:2005vu,Bhagwat:2006pu}
and baryons \cite{Oettel:2002wf,Alkofer:2004yf,Eichmann:2007nn,Alkofer:2009jk,Nicmorus:2010sd,Nicmorus:2010mc}. 
Improvements to this truncation have been considered in the past and studies in this direction are under way
\cite{Watson:2004jq,Watson:2004kd,Fischer:2008wy,Fischer:2009jm,Chang:2009zb,Chang:2010xs}.
What we discuss in the present work is most easily exemplified in a simple truncation, but becomes more
important --- and thus relevant --- with any kind of increasing numerical effort necessitated by either a more involved truncation or
the study of a system of more than two constituents.

The paper is organized as follows: in sec.~\ref{sec:structure} we collect the necessary formulae regarding covariant
bound-state equations as they are obtained in the DSE approach to QCD. Section \ref{sec:numerical}
details the discretization of the integrals and the general numerical setup. Section \ref{sec:methods} contains 
numerical solution strategies for both the homogeneous and inhomogeneous bound state equations. In sec.~\ref{sec:application}
we apply the methods described to solve the homogeneous and inhomogeneous BSE for pseudoscalar mesons in rainbow-ladder 
truncation and analyze the efficiency of the algorithms. 
Conclusions and an outlook indicating both immediate and further possible applications of the strategies described herein
are offered in sec.~\ref{sec:conclusions}.

All calculations are performed in Euclidean space.

\section{Structure of covariant bound-state equations\label{sec:structure}}
\begin{figure}
\begin{center}
\includegraphics[width=\columnwidth,clip=true]{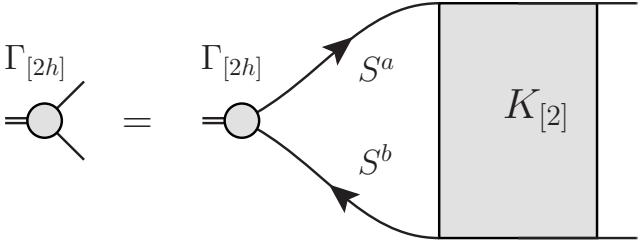}
\caption{\label{fig:hom_bse} The homogeneous BSE, Eq.~(\ref{eq:hom_bse})}
\end{center}
\end{figure}

\begin{figure}
\begin{center}
\includegraphics[width=\columnwidth,clip=true]{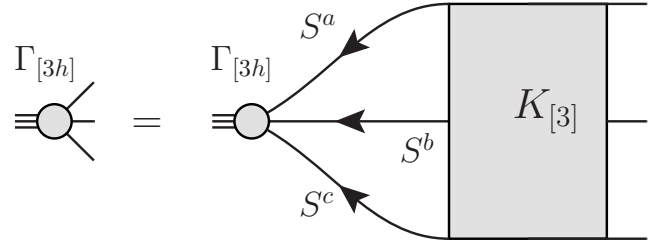}
\caption{\label{fig:faddeev} The homogeneous equation for a three-body bound state (covariant Faddeev equation), Eq.~(\ref{eq:faddeev})}
\end{center}
\end{figure}

In the DSE approach to QCD, mesons are described by general vertices connecting (anti-)quarks to objects carrying
the appropriate quantum numbers as demanded by the respective superselection rules. These vertices are the so-called
(inhomogeneous) Bethe-Salpeter amplitudes (BSAs), denoted by $\Gamma_{[2]}(k,P)$, which describes a two-particle system,
denoted by the subscript ${}_{[2]}$, with total
momentum $P$ and relative momentum $k$ of the constituents. The inhomogeneous BSA satisfies the inhomogeneous (vertex) BSE,
\begin{multline} \label{eq:inhom_bse}
\Gamma_{[2]} (k,P) = \Gamma_0(k,P)\\+\int_{q}\! K_{[2]}(k,q,P) S^a(q_+)\Gamma_{[2]} (q,P) S^b(q_-) \;,
\end{multline}
where $\Gamma^0(k,P)$ is a driving term with the quantum numbers of the system,
the Euclidean-space four-dimensional momentum integration is given by 
$\int_{q}=\int\frac{d^4\!q}{(2\pi)^4}\;$, $S^{a,b}(q_{\pm})$ denote the renormalized dressed (anti-)quark 
propagators, $K_{[2]}(k,q,P)$ represents the quark-antiquark
interaction kernel, and $q_{\pm}=q\pm\eta_{\pm}P$ are the (anti-)quark momenta with momentum partitioning $\eta_\pm$ such that
$\eta_++\eta_-=1$, respectively. The choice of the momentum partitioning is arbitrary and usually a matter of convenience,
e.g., for equal-mass constituents a convenient choice is $\eta_+=\eta_-=1/2$.
A graphical representation of Eq.~(\ref{eq:inhom_bse}) is given in Fig.~\ref{fig:inhom_bse}, where the arrows denote
dressed-quark propagators (analogously in Figs.~\ref{fig:hom_bse} and \ref{fig:faddeev}).

The solution of (\ref{eq:inhom_bse}), $\Gamma_{[2]}(P,k)$, contains both off-shell and on-shell information about the states
in a channel with the quantum numbers under consideration, which are fixed via the construction of $\Gamma^0(k,P)$ and $\Gamma_{[2]}(P,k)$.
In particular, $\Gamma_{[2]}(P,k)$ has poles\footnote{It should be noted that this formalism is also applicable to resonances,
where instead of a pole in $\Gamma_{[2]}(P,k)$ as a function of a real variable $P^2$ one expects a finite peak.} whenever the
total-momentum squared corresponds to the square of a bound-state mass in this channel (e.g.~\cite{Bhagwat:2007rj} and references therein).
If there exists a bound state and the corresponding on-shell condition, in Euclidean space $P^2=-M^2$, is met, the
properties of the bound state are described by the pole residues of Eq.~(\ref{eq:inhom_bse}). These residues, the homogeneous
BSAs $\Gamma_{[2h]}(k,P)$, can be obtained from the corresponding homogeneous BSE,
\beq  \label{eq:hom_bse}
\Gamma_{[2h]}(k,P)=\int_{q}\! K_{[2]}(k,q,P) S^a(q_+)\Gamma_{[2h]}(q,P) S^b(q_-) \;,
\eeq
depicted in Fig.~\ref{fig:hom_bse}.

For Baryons, an analogous construction can be made, where the homogeneous equation for the on-shell amplitude is a covariant
three-quark equation often referred to as a covariant Faddeev equation \cite{Eichmann:2009qa,Eichmann:2009fb,Eichmann:2009zx},
which may be written as
\begin{multline} \label{eq:faddeev}
\Gamma_{[3h]}(k_1,k_2,P) = \int_{q_1,q_2}\!\!\!\!S^a(p_1)S^b(p_2)S^c(p_3)\\ \times K_{[3]}(k_1,k_2,q_1,q_2,P) \Gamma_{[3h]}(q_1,q_2,P)\;,
\end{multline}
and a pictorial representation is given in Fig.~\ref{fig:faddeev}. Here, the kernel $K_{[3]}(k_1,k_2,q_1,q_2,P)$ subsumes all 
interactions of the three quarks with the individual momenta $p_i$, $i=1,2,3$, and the bound state is described by the
covariant three-quark on-shell amplitude $\Gamma_{[3h]}(k_1,k_2,P)$, which depends on the total momentum $P$ as well as two 
relative (Jacobi) momenta $k_1$ and $k_2$. Note that this equation contains an integral over two momenta, namely $q_1$ and $q_2$, 
thus inflating the size of the problem in terms of a numerical setup.

While in this work we will focus on the solution of bound-state equations such as (\ref{eq:inhom_bse}) or (\ref{eq:hom_bse}), a
note on the construction of the interaction kernels $K$ and the origin of the quark propagators $S$ as inputs in these equations
is in order. In QCD, in addition to the set of DSEs, the Green functions of the theory also satisfy Ward-Takahashi- and/or
Slavnov-Taylor identities, e.g.~\cite{Roberts:1994dr,Alkofer:2000wg}. These relate certain Green functions among each other
and provide guidance or even constraints in many cases, if one is to use a truncation and wants to make an Ansatz for,
say, $K_{[2]}$. For light-hadron physics, the axial-vector Ward-Takahashi identity is of particular interest, since it encodes
the chiral symmetry of QCD with massless quarks as well as its dynamical breaking (see, e.g.~\cite{Maris:1997tm,Maris:1997hd}
for details). In other words, satisfaction of this identity guarantees that the properties of the pion, the lightest hadron and
would-be Goldstone boson of dynamical chiral symmetry breaking follow the expected pattern, e.g.~the pion mass vanishes in
the chiral limit, and leads to a generalized Gell-Mann--Oakes--Renner relation valid for all pseudoscalar mesons
\cite{Maris:1997hd,Holl:2004fr}. The rainbow-ladder truncation of the DSE-BSE system satisfies the axial-vector Ward-Takahashi
identity. Beyond rainbow-ladder truncation, satisfaction of this identity can be achieved on more general terms and has been
implemented throughout light-hadron studies of the past years in this approach,
e.g.~\cite{Munczek:1994zz,Bender:1996bb,Watson:2004jq,Watson:2004kd,Bhagwat:2004hn,Fischer:2008wy,Fischer:2009jm,Chang:2009zb}.

More concretely, the satisfaction of this identity leads to a close relation of the interaction kernel 
$K_{[2]}$ in the quark-antiquark BSE and the quark self-energy present in the DSE of the quark propagator, 
the QCD gap equation. The consistent use of such related kernels in the BSE and the gap equation provides the 
proper input for the BSE in terms of the quark propagators
$S$ as solutions of the symmetry-preserving version of the gap equation. This consistency can be maintained also 
in numerical studies with high accuracy. In the following we always assume that the gap equation has 
been solved with the appropriate self-energy to match $K_{[2]}$ and that the resulting quark propagator $S$ is known numerically.

Note that in all bound state equations the real momentum-integration variables and the imaginary total on-shell 
momentum of the bound state are combined in the (anti-)quark momenta $q_\pm$ and $q_1$, $q_2$, $q_3$ to complex 
four-vectors. As a consequence, the arguments of the dressing functions in
the quark propagator become complex. In the meson BSE, for example, the dressing functions are needed in a 
parabolic region in the complex plane defined by $q_\pm^2$.
Over the past years reliable numerical approaches to this problem have been developed and the required computations 
are well under control (for more details see \cite{Fischer:2005en,Fischer:2008sp,Krassnigg:2008gd}).

In this way, with the specification of the truncation used, one both decides the structure of the interaction kernel and
obtains the quark propagator consistent with this kernel.

\section{Numerical representation\label{sec:numerical}}
The first step towards a numerical representation of Eqs.~(\ref{eq:inhom_bse}), (\ref{eq:hom_bse}), and (\ref{eq:faddeev})
is the analysis of the Lorentz and Dirac structure of the respective amplitudes. This structure is a result of the
particular representation properties of the state under consideration under the Lorentz group, including the state's parity
and spin. Therefore, the bound state amplitudes are decomposed into Lorentz-covariant parts $T_i$ and Lorentz-invariant parts $F^i$,
respectively, reading
\beq\label{eq:decomp}
\Gamma = \sum_{i=1}^N T_i \;F^i\;,
\eeq
where the number of terms $N$ as well as the tensor structure of the $T_i$ and $\Gamma$ depend on the quantum
numbers of the bound state and all arguments have been suppressed for simplicity. The $T_i$ are usually 
referred to as \emph{covariants}, whereas we call the $F^i$ the \emph{components}
of the amplitude $\Gamma$. The $T_i$ represent a spin-determined basis for the bound state, and one is --- to some
extent --- free to choose the details thereof.

To be more concrete, we consider the case of mesons in more
detail, where two spin-$1/2$ fermions (the quark and antiquark) are combined to a boson with total spin $J$.
As a result one obtains a $4\times 4$-matrix structure with the correct Lorentz-transformation properties
for a particle of spin $J$, see e.g.~\cite{Smith:1969az}.
Consequently one has, in addition to the total meson momentum $P^\mu$ and the relative momentum $q^\mu$ between the constituents,
another four vector $\gamma^\mu$ of Dirac matrices to construct the elements of the BSA. The various combinations of
the momenta and $\gamma^\mu$ correctly encode the angular momentum structures inside the meson, i.e.~the contributions
of (anti-)quark spin and orbital angular momentum.

A scalar meson BSA, for example contains all possible Lorentz-scalar combinations of the three four-vectors
$P$, $q$, and $\gamma$ (the actual construction is given below for a pseudoscalar meson in Eqs.~(\ref{eq:orthobasis1})-(\ref{eq:orthobasis4})).
For the moment we note that while $\Gamma$ and the $T_i$ in Eq.~(\ref{eq:decomp}) in general depend on the
three four-vectors $P$, $q$, and $\gamma$ as such (which we denote by semicolons between arguments of an expression),
the components, being Lorentz- and Dirac-scalars, can only depend on
scalar products of the momenta involved, i.e.~$P^2$, $q^2$, and $q\cdot P$ in the meson case. Thus, writing
arguments explicitly, Eq.~(\ref{eq:decomp}) reads
\beq\label{eq:decompexpl}
\Gamma (\gamma;q;P) = \sum_{i=1}^N T_i (\gamma;q;P)\;F^i (P^2,q^2,q\cdot P)\;,
\eeq
where again the tensor structure of the $T_i$ (and correspondingly $\Gamma$) was not denoted explicitly, since it
is irrelevant to the following argument.

In a numerical study, it is mostly advantageous to use a basis that is orthonormal, meaning the covariants satisfy
\beq \label{eq:covariants_norm}
\mathrm{Tr}\left(T_i\cdot T_j\right)=\delta_{i,j}\;,
\eeq
which also defines a generalized scalar product on the space of $4\times4$ matrices (any occurring Lorentz indices are
understood to be summed over here). Note that, for a set of covariants which is neither orthogonal nor normalized, the following step
is more involved and we detail it in \ref{sec:generaldirac}. If one uses the decomposition (\ref{eq:decomp}), the 
bound state equations (\ref{eq:inhom_bse}), (\ref{eq:hom_bse}), and (\ref{eq:faddeev}) can be rewritten as coupled integral 
equations of the components depending on the scalar products of the momenta via the corresponding projections on the basis $T_i$.

More concretely, we consider the integrand in, e.g. the (in{\nolinebreak}-{\nolinebreak}){\nolinebreak}homogeneous meson BSE,
\beq  \label{eq:int_term}
K_{[2]}(\gamma;k;q;P) \;S^a(\gamma;q;P)\;\Gamma_{[2]} (\gamma;q;P)\; S^b (\gamma;q;P)\;.
\eeq
The amplitude $\Gamma_{[2]} (\gamma;q;P)$ expanded in the chosen Dirac basis $T_j (\gamma;q;P)$ and the result
is projected on $T_i (\gamma;k;P)$. Doing so, one obtains a matrix structure in the space
of covariants, and Eq.~(\ref{eq:int_term}) can be written as a matrix-vector multiplication in this space 
involving the BSE \emph{kernel matrix} $K^i_j(k;q;P)$:
\begin{multline}\label{eq:kernelmatrix}
K^i_j(k;q;P) F^j(P^2,q^2,q\cdot P)=\\
\mathrm{Tr}\left[T_i(\gamma;k;P)\; K_{[2]}(\gamma;k;q;P)\;S^a(\gamma;q;P)\right.\\
\times\left.  T_j(\gamma;q;P)\; S^b(\gamma;q;P)\right]  F^j(P^2,q^2,q\cdot P)\;,
\end{multline}
where the sum over the repeated index $j$ is implied.

The index $j$ of the components $F^j(P^2,q^2,q\cdot P)$ can thus be viewed as a vector index, which has to be 
contracted with the corresponding index of the kernel matrix $K^i_j(k;q;P)$.
Note that this procedure, although exemplified here for the case of mesons, is completely general, i.e., it 
applies to baryons as well and is valid for any choice of the interaction kernel $K$.

The next step is to make the dependence on the continuous momentum variables $P^2$, $q^2$, and $q\cdot P$
numerically accessible. To achieve this, we apply the so-called Nystr\"om or quadrature method
(cf. \cite[Chap. 4]{Delves:1985aa}), which amounts to replacing an integral by a sum over suitable quadrature
weights and points and neglecting the error term. Applying this method discretizes the integration variables,
and consequently also the momentum dependence on the left hand side. The homogeneous and the inhomogeneous bound
state equations can then be written as matrix equations in the covariants and the discretized momenta and read
\beq
F^{i,\mathcal{P}}_{[h]} = K_{j,\mathcal{Q}}^{i,\mathcal{P}} F^{j,\mathcal{Q}}_{[h]}\label{eq:hom_index}
\eeq
in the homogeneous case, and
\beq
F^{i,\mathcal{P}}=F_0^{i,\mathcal{P}}+ K_{j,\mathcal{Q}}^{i,\mathcal{P}} F^{j,\mathcal{Q}}
\eeq
in the inhomogeneous case. The indices $i,\;j$ label the components, the multi-indices $\mathcal{P},\;\mathcal{Q}$
stand for all discretized momentum variables (summation over repeated indices is implied). The matrix
$\mathbf K = K_{j,\mathcal{Q}}^{i,\mathcal{P}}$ is the same in both equations, and subsumes the interaction kernel,
the dressed propagators of the constituents, the Dirac- and Lorentz structure, as well as the discretized integrations.
It is applied to a vector $F^{i,\mathcal{P}}$ representing the homogeneous or inhomogeneous bound state amplitude.

As an alternative to the Nystr\"om method, one can expand the momentum dependence of the components into suitable
sets of orthogonal functions, which can then be integrated. In this approach, the index $\mathcal{P}$ of the vector
$F^{i,\mathcal{P}}$ contains the coefficients of the expansion rather than the values of the components at certain
points in momentum space (for applications in the present context, see e.g.~\cite{Bhagwat:2002tx,Dorkin:2010ut}).

A partial application of this alternative is the use of a Chebyshev expansion of the dependence in an angle
variable as described in \ref{sec:integration}, where one only keeps a finite number of Chebyshev moments
in the representation of the amplitude. This step has been widely used in DSE studies of hadron
spectra and properties, and the fidelity of the approximation investigated in detail, see
e.g.~\cite{Maris:1997tm,Oettel:2001kd,Alkofer:2002bp}. While for studies of hadron masses a few moments are sufficient,
more are required in situations where considerable changes of the frame of reference are needed, such as form
factor calculations at large momentum transfer \cite{Holl:2005zi}; ultimately, in these situations the approximation
needs to be abandoned \cite{Maris:2005tt,Bhagwat:2006pu}.


\section{Solution methods\label{sec:methods}}
\subsection{Homogeneous equations: eigenvalue algorithms}
With the results of the preceding section, the homogeneous bound state equation (BSE or Faddeev equation), 
given in Eq.~(\ref{eq:hom_index}) in index notation, can be written as
\beq \label{eq:hom_bse_matvec}
\vec{F}_{[h]}=\mathbf{K}\cdot \vec{F}_{[h]}\;
\eeq
using matrix-vector notation.
As already mentioned in Sec.~\ref{sec:structure}, this equation is only valid at the on-shell points of the 
bound states in the respective channel, i.e.~at certain values of the
total momentum squared $P^2=-M_n^2$, where $n=0,1,2,\ldots$ numbers the ground- and all excited states in the channel.
To find such a value of $P^2$, one investigates the spectrum of $\mathbf{K}$ as a function of $P^2$, since Eq.~(\ref{eq:hom_bse_matvec})
corresponds to an eigenvalue equation (with the dependence on $P^2$ made explicit)
\beq \label{eq:hom_bse_matvec_lam}
\lambda(P^2)\vec{F}_{[h]}(P^2)=\mathbf{K}(P^2)\cdot \vec{F}_{[h]}(P^2)\;,
\eeq
where the eigenvalue $\lambda(P^2)=1$. In other words, to numerically approach a solution of the equation, a part of
the result has to be already known, namely the values $M_n^2$, or --- more precisely --- the mass of the state one
is looking for. The way out is a self-consistency argument, where the eigenvalue spectrum is plotted as a function of $P^2$
and those points with $\lambda_n(P^2)=1$ are identified: the largest eigenvalue determines the ground state, the smaller ones
in succession the excitations of the system (see also Fig.~\ref{fig:hom_solution} below). Typically, one is interested 
in roughly up to five eigenvalues, since higher excitations are both not well-enough understood in theory 
and hard to access experimentally.

A great variety of algorithms is available to numerically tackle these kinds of problems, and the most commonly 
used is a simple iterative method. Similar to the other algorithms discussed in this section, it relies on the 
multiplication of the matrix $\mathbf{K}$ on a vector and can successively be applied to find also excited states, 
by projecting on states already obtained, see e.g.~\cite{Krassnigg:2003wy}. This simple method, however, is not able to 
resolve pairs of complex conjugate eigenvalues, which may, for example, occur in the meson BSE \cite{Ahlig:1998qf}. In 
addition, the total number of required matrix-vector multiplications increases for every additional eigenvalue, as 
demonstrated in Sec.~\ref{sec:application}.

These difficulties are overcome by the use of more advanced algorithms. For this purpose, we use the 
implicitly restarted Arnoldi factorization \cite{Sorensen:1996aa}, which is frequently applied in lattice QCD studies, 
e.g.~\cite{Joergler:2007sh}. An application of this algorithm to bound state calculations is demonstrated in 
Sec.~\ref{sec:application}, where we use it to solve the pseudoscalar-meson BSE and compare the efficiency of both methods.

\subsection{Inhomogeneous equations: matrix inversion}
\label{sec:inhom_methods}
In the most compact notation, the inhomogeneous bound state equation can be written as
\beq \label{eq:inhom_bse_matvec}
\vec{F}(P^2)=\vec{F_0}(P^2)+\mathbf{K}(P^2)\cdot \vec{F}(P^2)
\eeq
where the matrix $\mathbf K(P^2)$ is identical to the one in Eq. (\ref{eq:hom_bse_matvec}), and the vector $\vec{F_0}$ is given
by the decomposition of $\Gamma_0$ according to Eq. (\ref{eq:decomp}), $\Gamma^0=\sum_i T_i F_0^i$ together with 
the discretization of a possible momentum dependence.

Again, the simplest method to treat this problem is a direct iteration. Mathematically, this corresponds to the representation of the
solution by a von Neumann series (cf. \cite[Chap. 4]{Delves:1985aa}), which can be shown to converge as long as the norm of the operator
$\mathbf K$ is smaller than one, $\|\mathbf K \| < 1$. For matrices, this norm can be related to the largest eigenvalue, such that for
$P^2>-M_0^2$, the iteration converges. When $P^2$ approaches the ground state position $-M^2_0$ from above, the number of iterations 
necessarily grows, and no convergence is obtained if $P^2\leq -M_0^2$, as demonstrated in Sec. \ref{sec:application}.

However, a solution is possible for any $P^2$ if one rewrites Eq.~(\ref{eq:inhom_bse_matvec}) as
\beq \label{eq:inhom_sol}
\vec{F} = (\mathbf 1 - \mathbf K)^{-1}\cdot \vec{F}_0 \;,
\eeq
i.e., $\vec{F}$ is given by the inhomogeneous term $\vec{F}_0$ multiplied by the matrix inverse of $(\mathbf 1 - \mathbf K)$.
$\vec{F}$ can then be computed by e.g.~inverting the matrix exactly, which has been successfully used to resolve 
bound-state poles in the inhomogeneous amplitude, as shown in \cite{Bhagwat:2007rj} in the case of mesons. On the downside, the direct
inversion of a matrix is computationally expensive, and it is not straightforward to parallelize the procedure.

A better approach is to view Eq.~(\ref{eq:inhom_sol}) as a linear system whose solution is to be found. Equations
like this are very common and several algorithms have been developed for their solution. In particular, if the
matrix $(\mathbf 1 - \mathbf K)$ is big, Eq. (\ref{eq:inhom_sol}) is a typical application for the so-called Conjugate
Gradient (CG) algorithms. Many types of these iterative Krylov-space methods are available. In the case of the bound-state equations
considered here, the matrices involved are neither hermitian nor symmetric, such that a good choice is the well-known
Bi-Conjugate-Gradients stabilized (BiCGstab) algorithm \cite{vanderVorst:1992aa}, which is widely used for example
in lattice QCD (cf.~\cite[Chap. 6.2]{Gattringer:2010zz}, where also the algorithm is presented in detail).

\section{Application: numerical solution of the meson BSE\label{sec:application}}

As an illustration, we apply the algorithms discussed above to solve the homogeneous and inhomogeneous
pseudoscalar-meson BSEs and compare their efficiency in terms of the number of matrix-vector multiplications needed to
achieve a specified accuracy. For bigger problems like baryons in a three-quark setup, the kernel matrix typically
does not fit into memory, and thus has to be recomputed on the fly in each iteration- or matrix-vector multiplication step.
In this case, one matrix-vector multiplication is rather time consuming and it is desirable to keep the number of necessary
multiplications as small as possible.

For our test case here, however, we study the pseudoscalar-meson BSE, where the kernel matrix is small, but one
can still investigate the questions at hand. We employ the rainbow-ladder truncation, i.e.~the
rainbow approximation in the quark propagator DSE together with a ladder truncation of the corresponding quark-antiquark BSE.
We define the difference in relative momenta $k-q=:\ell$ and the transverse projector with respect to the momentum $q$ as
$T^{\mu\nu}(q):=\left(\delta_{\mu\nu}-\frac{q_\mu q_\nu}{q^2}\right)$.
In this truncation, the kernel $K_{[2]}(\gamma;k;q;P)$ of the BSEs, Eqs.~(\ref{eq:inhom_bse}) and (\ref{eq:hom_bse}) is then given by
\begin{equation}\label{eq:rlkernel}
K_{[2]}(\gamma;k;q;P) = -\gamma_\mu  \frac{\mathcal D\left(\ell^2\right)}{\ell^2}T^{\mu\nu}(\ell) \gamma_\nu\;.
\end{equation}
where the effective interaction as a function of the momentum-squared $s$, introduced in Ref.~\cite{Maris:1999nt}, reads
\beq
\mathcal D(s) =  D\; \left(\frac{4 \pi^2}{{\omega}^6} s e^{-s/{\omega}^2}\right) + \mathcal F_{UV}(s)\;.
\eeq
The term $\mathcal F_{UV}(s)$ implements the perturbative running coupling of QCD for large $s$,
preserving the one-loop renormalization-group behavior of QCD. The Gaussian term models the enhancement in the 
intermediate-momentum regime necessary to produce a reasonable amount of dynamical chiral symmetry breaking. It contains the parameters
of the model, $D$ and $\omega$, describing the overall strength and momentum-space width (corresponding to an
inverse effective range) of the interaction. The behavior of the effective interaction in the far infrared is 
expected to be of minor relevance to ground-state properties (cf. \cite{Blank:2010pa} and references therein). 
For the present study, we make a common choice for the parameters, namely $D=0.93\,\mathrm{GeV}^2$ and 
$\omega=0.4\,\mathrm{GeV}$ (for full details on the truncation, the effective coupling, or the
effects of other parameter values, see e.g.~\cite{Maris:1997tm,Maris:1999nt,Krassnigg:2009zh}).

\subsection{Kernel setup}
The above definitions together with the dressed quark propagators computed already completely specify the ingredients
of the BSE. We investigate light quarks in analogy to \cite{Krassnigg:2009zh}, where the current-quark mass in an isospin-symmetric
setup was adjusted to fit the bound-state mass of the $\rho$ meson. The details of the discretization of the kernel matrix proceed
as follows: The orthonormal pseudoscalar covariants, constructed to satisfy Eq.~(\ref{eq:covariants_norm}), read
\begin{eqnarray}\label{eq:orthobasis1}
 T_1 &=&\frac{\gamma_5}{2},\;\\
 T_2 &=&\frac{\gamma_5 (\gamma\cdot P)}{2 \sqrt{-P^2}},\;\\
 T_3 &=&\frac{\gamma_5 ( (\gamma\cdot q)- \frac{(\gamma \cdot P) (P\cdot q)}{P^2})}{ 2 \sqrt{\frac{(P\cdot q)^2}{P^2}-q^2}},\;\\
 T_4 &=&\frac{\frac{1}{2} i \gamma_5((\gamma\cdot q) (\gamma\cdot P)  - (
 \gamma\cdot P) (\gamma\cdot q))}{2 \sqrt{P^2 q^2-(P\cdot q)^2}}\label{eq:orthobasis4} \;.
\end{eqnarray}
Choosing the rest frame of the quark-anti-quark system, and applying the parametrization and discretization 
as described in \ref{sec:integration}, the kernel matrix Eq.~(\ref{eq:kernelmatrix}) in our setup becomes
\begin{multline}
\mathbf K = K^{i,r,s}_{j,l,m}(P) = \\ -\frac{4}{3 (2 \pi)^3} w[q_l^2] w[z_m] \int_{-1}^1 \!\! dy\; 
\frac{\mathcal D\left(\ell^2\right)}{\ell^2}T^{\mu\nu}(\ell) \\
\times\mathrm{Tr}\left[
T_i(\gamma;k;P) \gamma_\mu 
S(q_+)T_j(\gamma;q;P) S(q_-) \gamma_\nu
\right]\;,
\end{multline}
where $w[q_l^2]$, $w[z_m]$ denote the quadrature weights and the replacements $k^2\rightarrow k_r^2$, 
$z_k\rightarrow z_s$, $q^2\rightarrow k_l^2$, $z\rightarrow z_m$ have been made in all occurring momenta to 
implement the discretization. Therefore, the indices $i;j$ label the components and $r,s;l,m$ the momentum 
space points. For the following calculations, we use $N_q=32$ and $N_z=24$, such that $\mathbf K$ has the 
dimensions $(32,24,4)\times(32,24,4)$.
\begin{figure}
\begin{center}
\includegraphics[width=\columnwidth,clip=true]{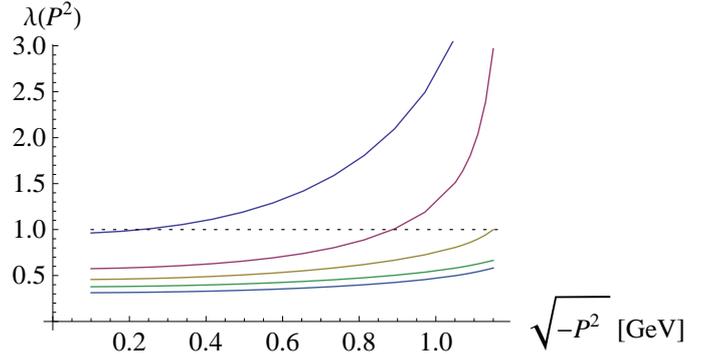}
\caption{\label{fig:hom_solution}The five largest eigenvalues of the homogeneous BSE plotted over $\sqrt{-P^2}$
which corresponds to the bound state mass $M$ where $\lambda=1$ indicated by the horizontal dashed line.
The ground-state (leftmost intersection) solution vector has positive $C$-parity (pion), the second
has negative (exotic) and the third again has positive $C$-parity (excited pion).}
\end{center}
\end{figure}

\begin{figure}
\begin{center}
\includegraphics[width=\columnwidth,clip=true]{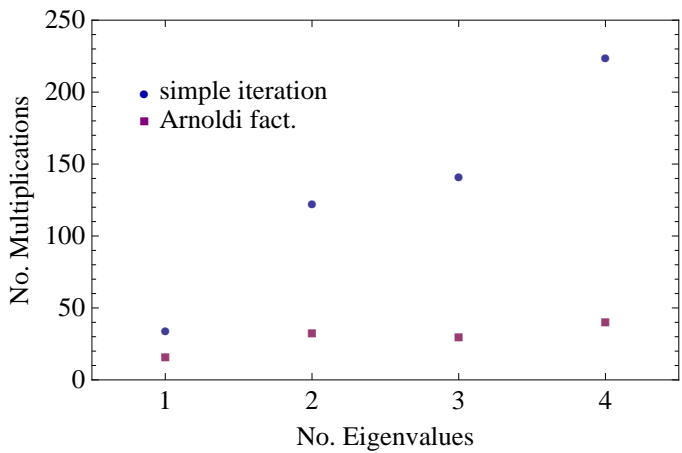}
\caption{\label{fig:hom_compare} The number of matrix-vector multiplications needed for convergence of the
simple iteration (circles) and the Arnoldi factorization (squares), plotted against the number of eigenvalues computed at
a (typical) fixed value of $P^2=-M_0^2$.}
\end{center}
\end{figure}

\subsection{Homogeneous BSE}
To solve the homogeneous BSE, we use both the MPI based version of the ARPACK library (an implementation of the
implicitly restarted Arnoldi factorization) and the simple iteration. Fig.~\ref{fig:hom_solution} shows the largest
five eigenvalues of $\mathbf K$ for our example. Bound-state masses can be identified by the positions at which an
eigenvalue curve crosses one (dashed line in the figure) and from left to right correspond to the ground- and first-excited,
second-excited, etc., states. Note that in our approach we do not restrict the symmetry of the amplitudes
(eigenvectors) with respect to the the angular variable $z$, such that we obtain homogeneous solutions of both positive
and negative charge-conjugation parity ($C$-parity) for equal-mass constituents and a choice of $1/2$ for the
momentum-partitioning parameters $\eta_\pm$, as indicated above (for more details on the definition and calculation
of $C$, see \ref{sec:cparity}). In the pseudoscalar case, a negative $C$-parity is considered \emph{exotic},
since it is not available for a $\bar{q}q$ state in quantum mechanics. However, it appears naturally in a quark-antiquark
BSE setup, where our main interest here comes from a systematic point of view. A more general discussion of states with exotic
$C$-parity in this formalism and their possible interpretations can be found, e.g., in 
\cite{Smith:1969az,Nakanishi:1969ph,Ahlig:1998qf,Burden:2002ps}.

To compare the efficiency of the two algorithms, Arnoldi factorization versus iteration, we compute the one to
four largest eigenvalues of $\mathbf K$ and compare the convergence in terms of the number of iterations needed 
to obtain an absolute accuracy of the eigenvector of $\epsilon = 10^{-8}$, at a typical value of 
$P^2=-M_0^2=0.0527\,\mathrm{GeV}^2$. The results, given in Fig.~\ref{fig:hom_compare}, show that for the first 
eigenvalue the Arnoldi factorization needs only half as many matrix-vector multiplications as the simple iteration. 
With increasing number of eigenvalues, the use of this advanced algorithm becomes even more advantageous.
 
Another interesting observation from Fig.~\ref{fig:hom_compare} is that the Arnoldi factorization was more
efficient for three eigenvalues than when only two were requested. This is most likely due to a ``clustering''
of eigenvalues number two and three for the algorithm, an effect which appears for eigenvalues close together
and is also related to the eigenvectors. In this particular case, eigenvectors two and tree have opposite $C$-parity
or $z$-symmetry, which may make them more easily distinguishable for the algorithm and more easy to obtain as 
a result. The ARPACK library is very efficient at evaluating all eigenvalues in such a cluster, while convergence
is slower, if one asks for only one or a few of the eigenvalues in the cluster (see also the ARPACK users guide \cite{Lehoucq:1998aa}).

\begin{figure}
\begin{center}
\includegraphics[width=\columnwidth,clip=true]{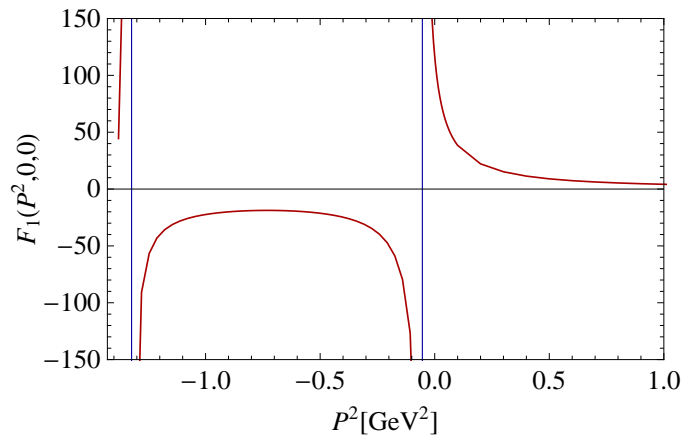}
\caption{Component $F_1(P^2,0,0)$ of the inhomogeneous pseudoscalar amplitude calculated using BiCGstab
vs. the square of the total momentum $P^2$. The vertical lines mark the pole positions, corresponding
to the pion ground- and first excited state ($J^{PC}=0^{+-}$).\label{fig:inhom_solutions}}
\end{center}
\end{figure}

\subsection{Inhomogeneous BSE}
We apply the direct iteration (summation of the von Neumann series) and the inversion using the BiCGstab algorithm
to solve the inhomogeneous BSE (\ref{eq:inhom_bse}), in the setup described above for pseudoscalar quantum numbers.

\begin{figure}
\begin{center}
\includegraphics[width=\columnwidth,clip=true]{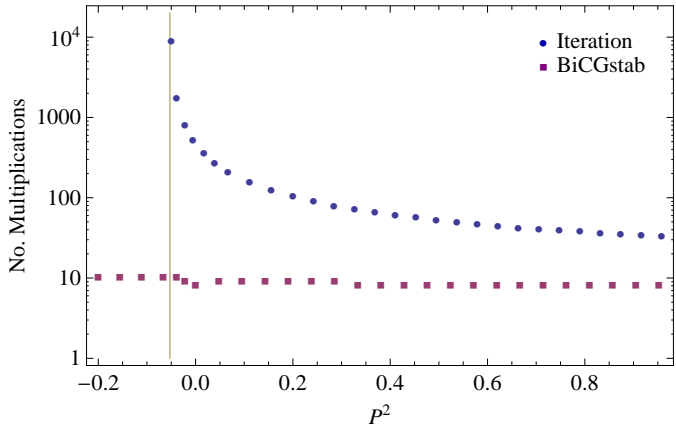}
\caption{\label{fig:inhom_compare} The number of matrix-vector multiplications needed for convergence
of the iterative solution of Eq.~(\ref{eq:inhom_bse}) (circles) and the BiCGstab algorithm (squares), plotted on a logarithmic
scale against the square of the total momentum $P^2$ of the amplitude. The vertical line indicates the position of the
ground state $P^2=-M_0^2$ of the system. Note that the straightforward iteration does not converge for $P^2\leq-M_0^2$.}
\end{center}
\end{figure}

In the inhomogeneous case not only the structure of the amplitude determines the quantum numbers of the solution
but also that of the inhomogeneous term $\Gamma_0$. Following \cite{Maris:1997hd}, a possible choice for pseudoscalars is
\beq \label{eq:gamma0}
\Gamma_0=Z_4\gamma_5\;,
\eeq
where $Z_4$ is a renormalization constant obtained from the gap equation (cf.~\cite{Maris:1997tm}). With this
choice (pseudoscalar, positive $C$-parity), no poles corresponding to negative $C$-parity appear in the solution,
as can be seen from Fig.~\ref{fig:inhom_solutions}. The curve shown in this figure 
has been obtained with the BiCGstab algorithm, because as described in Sec.~\ref{sec:inhom_methods} the direct iteration 
fails to converge if $P^2\leq-M_0^2$. This is demonstrated in Fig.~\ref{fig:inhom_compare}, where the number 
of matrix-vector multiplications needed for convergence is plotted against $P^2$ for both methods.

It is clear that the number of matrix-vector multiplications needed for the direct iteration diverges as $P^2$
approaches $-M_0^2$ (note that Fig.~\ref{fig:inhom_compare} uses a logarithmic scale on the vertical axis). The inversion with BiCGstab,
however, converges for all $P^2$ with nearly the same speed, needing approximately 10 matrix-vector multiplications.

\section{Conclusion and Outlook\label{sec:conclusions}}
We have investigated aspects and benefits of different numerical solution methods for the
pseudoscalar meson BSE in a rainbow-ladder truncation of the DSE approach to QCD. In a realistic
model, our comparison was aimed at a small-scale test and subsequent identification of efficient algorithms
for the numerical approach to bound-state problems in this covariant continuum approach to a quantum field
theory. Since the algorithms used and advocated are applicable in a very general way to the matrix
representations of the integral kernels appearing in such bound-state studies, the strategies proposed here
are valuable for similar studies of bound states in more involved truncations than rainbow-ladder on one hand.

On the other hand, any bound-state problem involving more than two constituents, starting with but not
limited to baryons in a three-quark setup, needs efficient methods to perform sophisticated numerical studies,
given the computational resources presently available. In general one will find that in both examples given here
one simply has to deal with a larger kernel matrix and thus efficiency of the algorithms used is of the essence.

One particularly interesting observation with regard to present-day covariant three-quark studies is that computing
bound-state masses is more efficient using the inhomogeneous equation than the homogeneous one. In addition, the 
kernel matrices often need to be constructed on the fly due to limited system memory. With these two aspects 
combined, such a problem appears to be an ideal application for the field of GPU computing.



\appendix
\section{Aspects of four-dimensional momentum integration\label{sec:integration}}
We use 4-dim. spherical coordinates, such that the momentum integration is written as
\beq
\int_{-\infty}^\infty \!\! d^4q\rightarrow \int_0^\infty \!\! d(q^2) \frac{q^2}{2}\int_{-1}^1 \!\! 
dz \sqrt{1-z^2}\int_{-1}^1\!\! dy \int_0^{2 \pi} \!\! d\phi\;.
\eeq
In the meson BSE, there are three relevant momenta: $P$ (total momentum), $k$ (relative momentum), 
and $q$ (loop momentum). Subsequently, we choose $P$ to be in the rest-frame of the bound state,
\beq\label{eq:rf1}
P=\left(0,0,0,\sqrt{P^2}\right)\;.
\eeq
The other momenta are chosen accordingly and read
\beq 
k = \sqrt{k^2}\left(0,0,\sqrt{1-z_k^2},z_k \right)
\eeq
and
\beq 
q = \sqrt{q^2}\left(0,\sqrt{1-z^2}\sqrt{1-y^2},\sqrt{1-z^2}y ,z \right)\;.\label{eq:rf2} 
\eeq
In this parametrization, the integration $\int d\phi$ is trivial.

The components of the amplitude on the left hand side of the BSEs Eqs.~(\ref{eq:inhom_bse}) 
and (\ref{eq:hom_bse}) depend on the scalar products $P\cdot k=z_k \sqrt{k^2} \sqrt{P^2}$, 
$P^2$, and $k^2$. Inside the integral, on the right hand side, the scalar products become 
$P\cdot q= z\sqrt{q^2} \sqrt{P^2}$, $P^2$, and $q^2$. Thus, the BSE kernel matrix $\mathbf K$ 
induces the following mapping on the momentum variables
\beq 
(q^2,z) \mapsto (k^2,z_k)\;, 
\eeq
such that the integration $\int dy$ does not add a dimension $\mathbf K$, although it is not trivial.

After choosing a parametrization of the momentum integration, the next step is to discretize 
the momentum dependence. In this work, we straightforwardly apply the quadrature method, and replace
\beq 
\int_0^\infty \!\! d(q^2) \frac{q^2}{2}\int_{-1}^1 \!\! dz \sqrt{1-z^2}\rightarrow \sum_{l=1}^{N_q} \sum_{m=1}^{N_z} w[q_l^2] w[z_m]\;, 
\eeq
where $w[q_l^2]$, $w[z_m^2]$ denote the quadrature weights and $q_l^2$, $z_m$ the corresponding nodes. 
The factors of $q^2/2$ and $\sqrt{1-z^2}$ have been absorbed in the weights.

Note that, especially for the integration over $z$, it is advantageous to use a quadrature rule 
whose weights include $\sqrt{1-z^2}$ by construction, e.g., the Gauss-Chebyshev type 2 rule.

An alternative, advantageous and widely used in the calculations of hadron spectra is to apply the 
quadrature method discussed above only to $\int d(q^2)$ and to resolve the $z$-dependence and by an 
expansion in Chebyshev polynomials of the second kind. The components are then written as
\beq 
F^i(P^2,q^2,z)=\sum_{m=1}^M {}^mF_i(q^2,P^2)U_m(z)\;, 
\eeq
with ${}^mF_i(q^2,P^2)$ the so-called Chebyshev \emph{moments}, which retain only the functional 
dependence on $k^2$ and $P^2$. The number of terms $M$ taken into account is finite in practice, but infinite
in principle. The Chebyshev polynomials of the second kind $U_m(z)$ satisfy the orthogonality relation
\beq \label{eq:cheby_ortho}
\frac{2}{\pi}\int dz\sqrt{1-z^2}\;U_m(z)\,U_n(z)=\delta_{mn}\;.  
\eeq
To obtain a matrix structure not only in the covariants, but also in the Chebyshev moments, the 
above expansion is inserted in Eq.~(\ref{eq:kernelmatrix}), and is then projected on one moment by 
use of Eq.~(\ref{eq:cheby_ortho}). This finally leads to
\begin{multline}
\int_q K^{i,m}_{j,n}(k;q;P)\;{}^nF_j(q^2,P^2)=\\
\left(\,\frac{2}{\pi}\int_q \int_{-1}^1 \!\! dz_k\,U_m(z_k)\,\mathrm{Tr}\left[T_i(\gamma;k;P)\; K_{[2]}(\gamma;k;q;P) \right.\right.\times\\
 \left.\left. S^a(\gamma;q;P)\; T_j(\gamma;q;P)\; S^b(\gamma;q;P)\right]\,U_n(z)\,\right)\;{}^nF_j(q^2,P^2) \;.
\end{multline}
Again, the sum over the repeated indices $j,\;n$ is implied.

\section{(Generalized) $C$-parity\label{sec:cparity}}
The action of the $C$-parity transformation on the meson BSA is defined via
\beq
\overline{\Gamma}(P;k)=(C \Gamma(P;-k)C^{-1})^t\;,
\eeq
where $t$ denotes the matrix-transpose, and the charge-conjugation matrix $C=\gamma_2\gamma_4$. 
If the $C$-parity is a good quantum number of the system, $\Gamma(P;k)$ is an eigenstate of the $C$-parity 
operation given above, with eigenvalues $\lambda_C=\pm1$,
\beq\label{eq:cparity_eval}
(C \Gamma(P;-k)C^{-1})^t=\lambda_C \Gamma(P;k)\;.
\eeq
The amplitudes occurring in Eq.~(\ref{eq:cparity_eval}) can be decomposed into covariants and components 
according to Eq.~(\ref{eq:decompexpl}), such that in the rest frame of the bound state with the notation 
introduced in Eqs.~(\ref{eq:rf1})~-~(\ref{eq:rf2}),
\begin{multline}
\sum_{i=1}^{N}(C\,T_i(P;-k)\,C^{-1})^t\,F^i(P^2,k^2,-z_k)\\
 = \lambda_C \sum_{i=1}^{N}T_i(P;k)F^i(P^2,k^2,z_k)\;.
\end{multline}
Projecting this equation on one covariant using the orthogonality relation (\ref{eq:covariants_norm}), we obtain
\begin{multline}
\bar{F}^j(P^2,k^2,z_k):=\\
\sum_{i=1}^{N}\mathrm{Tr}\left[T_j(P;k) (C\,T_i(P;-k)\,C^{-1})^t\right]\,F^i(P^2,k^2,-z_k)\\
 = \lambda_C F^j(P^2,k^2,z_k)\;.
\end{multline}
The next step is to discretize the momenta using the quadrature method, such that the functional 
dependence of the component vectors on the momentum variables can be represented in index notation,
\begin{eqnarray}
F^j(P^2,k^2,z_k)&\Rightarrow& F^{j,l,m}\;,\\
\bar{F}^i(P^2,k^2,z_k)&\Rightarrow& \bar{F}^{i,l,m}\;.
\end{eqnarray}
Now, the amplitudes are given as complex vectors, such that the canonical scalar product on 
$\mathbf C^n$ can be used to solve for $\lambda_C$,
\beq\label{eq:cparity}
\lambda_C=\frac{\bar{F}^{j,l,m}(F^{j,l,m})^\ast}{F^{i,r,s}(F^{i,r,s})^\ast}\;,
\eeq
where $\ast$ denotes complex conjugation, and repeated indices are summed over.

For states with definite $C$-parity $\lambda_C=\pm 1$. As can be seen from the above equations, 
the $C$-parity is determined by the (anti-)symmetry of the amplitudes with respect to $z_k$. For 
states which are not eigenstates of the $C$ operation, Eq.~(\ref{eq:cparity}) can be used to \emph{define} a 
"generalized $C$-parity". It lies between $-1$ and $1$, and its deviation from these values indicates the 
asymmetry of the state caused, e.g., by mass difference of the constituents.

It is interesting to note that the $C$-parity as a symmetry property of the eigenvectors 
of $\mathbf K$ is constant over the whole range of $P^2$, even if the on-shell condition 
for this state is not fulfilled, i.e. $P^2\neq-M^2$.

\section{Using a non-orthogonal Dirac basis\label{sec:generaldirac}}
Consider the inhomogeneous BSE written in the general form
\begin{multline}\label{generalbse}
f(\gamma;k;P)\Gamma(\gamma;k;P)=Z(\gamma;k;P)\\
- \int_q\;K_1(\gamma;k;q;P)\Gamma(\gamma;q;P)K_2(\gamma;k;q;P)\;,
\end{multline}
where the dependence of every term on all variables including $\gamma$ matrices is given explicitly. The ;
between variables again denotes a dependence on complete four-vectors. $K_1$ and $K_2$ represent generalized
formal kernel pieces, $Z$ a general driving term, and $f$ an arbitrary function of its arguments. To transform this equation into a set
of coupled integral equations for components and then use Chebyshev moments, which are described in \ref{sec:integration},
we write the BSA as the sum over its covariants $T_i$ and Lorentz- as well as Dirac-scalar components $F^i$ and the latter
as sums over Chebyshev polynomials and moments
\begin{equation}
\Gamma(\gamma;k;P)=\sum_{i=1}^M\sum_{j=1}^{N}T_j(\gamma;k;P)\;{}^iF_j(k^2,P^2)U_i(z_k)\;,
\end{equation}
where $M$ is the number of Chebyshev polynomials taken into account and $N$ is the number of covariants in the BSA.
The Chebyshev polynomial of the second kind $U_i(z_k)$ depends on $z_k:=k\cdot P/\sqrt{k^2P^2}$.
Now we apply $T_n(\gamma;k;P)$ on Eq.~(\ref{generalbse}) from the left and take the Dirac trace.
The result is
\begin{multline}
\sum_{i=1}^M\sum_{j=1}^{N}A_{nj}(k^2,P^2,z_k){}^iF_j(k^2,P^2)U_i(z_k)=\\
Z_n(k^2,P^2,z_k) -\sum_{l=1}^M\sum_{m=1}^{N}\int_q\;\mathrm{Tr}\left[T_n(\gamma;k;P)\right.\\
\times K_1(\gamma;k;q;P)T_m(\gamma;q;P)\;{}^lF_m(q^2,P^2)\\
\left. \times U_l(z_q)K_2(\gamma;k;q;P)\right]\;,
\end{multline}
where
\begin{multline}
A_{nj}(k^2,P^2,z_k):=\mathrm{Tr}\left[T_n(\gamma;k;P)f(\gamma;k;P)T_j(\gamma;k;P)\right]\\
Z_n(k^2,P^2,z_k):=\mathrm{Tr}\left[T_n(\gamma;k;P)Z(\gamma;k;P)\right]
\end{multline}
The next step is to invert the matrix $A_{nj}$ for each set of coordinates $(k^2,P^2,z_k)$, and apply its
inverse to the equation, i.\,e., $\sum_{n=1}^{N}A^{-1}_{rn}$ from the left:
\begin{multline}
\sum_{i=1}^M{}^iF_r(k^2,P^2)U_i(z_k)=
\sum_{n=1}^{N}A^{-1}_{rn}(k^2,P^2,z_k)Z_n(k^2,P^2,z_k)\\
-\sum_{n=1}^{N}\sum_{l=1}^M\sum_{m=1}^{N}\int_q 
A^{-1}_{rn}(k^2,P^2,z_k)\mathrm{Tr}\left[T_n(\gamma;k;P)K_1(\gamma;k;q;P)\right.\\
\times \left. T_m(\gamma;q;P)\;
{}^lF_m(q^2,P^2)U_l(z_q)K_2(\gamma;k;q;P)\right]\;\quad\mbox{}
\end{multline}
The last step is the projection with the help of Chebyshev polynomials via
$\frac{2}{\pi}\int dz_k\sqrt{1-z_k^2}\;U_j(z_k)$ from the left (cf.~Eq.~(\ref{eq:cheby_ortho})) and one obtains
\begin{multline}
{}^jF_r(k^2,P^2)=V_Z(j,r,k^2)-\sum_{l=1}^M\sum_{n,m=1}^{N}\frac{2}{\pi}\int_q\int_{z_k}\sqrt{1-z_k^2}\;\\
\times U_j(z_k) A^{-1}_{rn}(k^2,P^2,z_k)
\mathrm{Tr}\left[T_n(\gamma;k;P)K_1(\gamma;k;q;P)\right.\\
\left. \times T_m(\gamma;q;P)\;
{}^lF_m(q^2,P^2)U_l(z_q)K_2(\gamma;k;q;P)\right]\quad\mbox{}
\end{multline}
with the driving term given by
\begin{multline}
V_Z(j,r,k^2):=\frac{2}{\pi}\sum_{n=1}^{N}\int dz_k\sqrt{1-z_k^2}\;U_j(z_k)\\
\times A^{-1}_{rn}(k^2,P^2,z_k)Z_n(k^2,P^2,z_k)\;.
\end{multline}
This procedure does not require the set of covariants to be orthogonal, it is completely general, also with respect to
the kernel, the driving term, and possible terms multiplying the amplitude on the left-hand side of the BSE.
All terms and projections are included correctly via the matrix $A$. In the case considered in the present work, $f=1$
and the driving term has the standard form for pseudoscalar mesons, Eq.~(\ref{eq:gamma0}).

\section*{Acknowledgments}
We would like to acknowledge valuable discussions with R.~Alkofer, C.\,S.~Fischer, C.~Gattringer, C.\,B.~Lang, A.~Maas, D.~Mohler, 
C.\,D.~Roberts, M.~Schwinzerl, and S.\,V.~Wright.
This work was supported by the Austrian Science Fund \emph{FWF} under project no.\ P20496-N16,  and
was performed in association with and supported in part by the \emph{FWF} doctoral program no.\ W1203-N08.

\end{document}